# Strong Effects of Interlayer Interaction on Valence-Band Splitting in Transition Metal Dichalcogenides

Garrett Benson[1], Viviane Zurdo Costa[1], Neal Border[1], Kentaro Yumigeta[2], Mark Blei[2], Sefaattin Tongay[2], K. Watanabe,[3] T. Taniguchi,[3] Andrew Ichimura[4], Santosh KC[5], Taha Salavati-fard,[6] Bin Wang,[6] and Akm Newaz[1]

[1]Department of Physics and Astronomy, San Francisco State University, San Francisco, California 94132, USA
[2] School for Engineering of Matter, Transport and Energy, Arizona State University, Tempe, Arizona 85287, USA
[3] National Institute for Materials Science, Namiki 1-1, Tsukuba, Ibaraki 305-0044, Japan
[4]Department of Chemistry and Biochemistry, San Francisco State University, San Francisco, California 94132, USA
[5]Chemical and Materials Engineering, San Jose State University, San Jose, California 95112, USA
[6]School of Chemical, Biological and Materials Engineering, University of Oklahoma, Norman, Oklahoma 73019, United States

Date (December 8th, 2021)

Abstract

Understanding the origin of valence band maxima (VBM) splitting in transition metal dichalcogenides (TMDs) is important because it governs the unique spin and valley physics in monolayer and multilayer TMDs. In this work, we present our systematic study of VBM splitting ($\Delta$) in atomically thin $MoS_2$ and $WS_2$ by employing photocurrent spectroscopy as we change the temperature and the layer numbers. We found that VBM splitting in monolayer $MoS_2$ and $WS_2$ depends strongly on temperature, which contradicts the theory that spin-orbit coupling solely determines the VBM splitting in monolayer TMDs. We also found that the rate of change of VBM splitting with respect to temperature ($m = \frac{\partial \Delta}{\partial T}$) is the highest for monolayer (-0.14 meV/K for $MoS_2$) and the rate decreases as the layer number increases ($m \sim 0$ meV/K for 5 layers $MoS_2$). We performed density functional theory (DFT) and the GW with Bethe-Salpeter Equation (GW-BSE) calculations to determine the electronic band structure and optical absorption for a bilayer $MoS_2$ with different interlayer separations. Our simulations agree with the experimental observations and demonstrate that the temperature dependence of VBM splitting in atomically thin monolayer and multilayer TMDs originates from the changes in the interlayer coupling strength between the neighboring layers. By studying two different types of TMDs and many different layer thicknesses, we also demonstrate that VBM splitting also depends on the layer numbers and type of transition metals. Our study will help understand the role spin-orbit coupling and interlayer interaction play in determining the VBM splitting in quantum materials and develop next-generation devices based on spin-orbit interactions.

Transition metal dichalcogenides (TMDs) exhibit remarkable properties resulting from their reduced dimensionality and crystal symmetry.[1-6] Monolayer TMDs (1L-TMDs) have a direct bandgap at the K point of the Brillouin zone, which is different from its corresponding few-layer and bulk counterparts with indirect band gaps.[1-6] The valence band maxima (VBM) at K point in TMDs split, which is attributed to the large spin-orbit coupling (SOC) in 1L-TMDs and the mixing of SOC and interlayer coupling for multilayer TMDs.[7-13] Strong SOC in TMDs originate from the $d-$orbitals of the heavy transition-metal atoms.[7] Unique crystal symmetry of TMDs causes coupling of spin and valley degrees of freedom.[14] The crystal symmetry of TMDs and SOC cause a sizable split ($\Delta > 100$ meV) in the valence band at the K point in the Brillouin Zone (BZ). This large splitting governs many observable physical phenomena such as the spin-Hall effect,[15] valley-Hall effect,[16] and optical circular dichroism.[17]

Despite the importance of the VBM splitting that causes the unique spin and valley physics of TMDs, the temperature and layer dependence of VBM has not been well studied experimentally. In this work, we present our systematic study of VBM splitting in $MoS_2$ and $WS_2$ as we change the temperature and the layer number by employing photocurrent spectroscopy. Surprisingly, we have observed that the VBM in 1L-TMDs depends on the temperature, which contradicts the theory that VBM splitting in 1L-TMDs originates solely from the SOC. Our finding also contradicts two previously reported experimental study that concluded that the VBM splitting in monolayer originates from SOC only.[13, 18] Intriguingly, we found that the temperature dependence is the strongest for 1L-$MoS_2$ than multilayer $MoS_2$. We also found that VBM in TMDs ($MoS_2$ and $WS_2$) also depends on the layer numbers and type of transition metals. We calculated the electronic band structure of bilayer $MoS_2$ by using density functional theory (DFT) and determined the absorption spectra using the GW-BSE approach, which agrees with experimentally measured VBM properties in $MoS_2$.

**RESULTS AND DISCUSSION**

To probe VBM splitting in atomically thin semiconductors, we have prepared molybdenum disulfide ($MoS_2$) and tungsten disulfide ($WS_2$) flakes that are electrically connected for transport measurement. All the samples reported here are covered by hexagonal boron nitride (hBN) for environmental protection. The $MoS_2$ and $WS_2$ flakes were mechanically exfoliated from bulk crystals onto a heavily doped silicon substrate capped with a 90 nm thick thermally grown $SiO_2$ film. The $MoS_2$ flakes were obtained from naturally grown rock and $WS_2$ samples were grown by chemical vapor transport (CVT) technique. The number of TMD layers was characterized by using optical microscopy, Raman spectroscopy, and atomic force microscopy (AFM).

For encapsulation with a few layers hBN, we prepare hBN flakes on $SiO_2$/Si substrate, which was picked by polyethylene terephthalate (PET). We used PET stamp to pick up the top hBN flake, atomically thin TMDs in sequence with accurate alignment using an optical microscope. The hBN/TMDs heterostructure was then stamped on a pre-fabricated Au electrode (70nm Au/ 5nm Cr) on a glass substrate (see Methods for details). The patterned Au electrodes were fabricated using optical lithography followed by thermal evaporation of metals. All devices were prepared on a glass substrate to avoid the photogating effect.[19-21]

Figure 1.a shows the optical images of a monolayer $MoS_2$ device encapsulated or covered by a hBN layer. The vertical design of the heterostructure sample is shown schematically in Fig.1a inset. Fig.1b shows the Raman spectrum for 1L, 2L, and 4L-$MoS_2$ samples. Confocal micro-Raman measurements were performed using commercial equipment (Horiba LabRAM 419 Evolution) (see Methods for details).

Reduced screening and strong light-matter interactions in van der Waals (vdWs) materials cause the formation of a wide range of many-body excitonic states. Intralayer excitons are bound electron-hole (*e-h*) pairs in the same layers, such as A-/B- excitons in TMDs originating from the split VBM at the K point. Hence, the measurement of A and B excitons provide a direct pathway to measure the VBM splitting. In addition to these many-body bound states, the band structure of 1L-TMDs causes a singularity in the joint density of states (JDoS) resulting in a unique exciton type, known as van Hove singularity excitons, also known as C and D excitons.[19, 22]

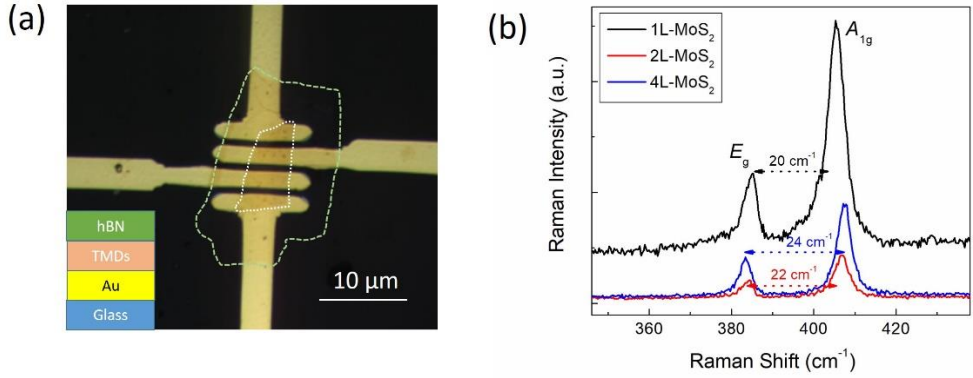

Figure 1: Left: Optical image of a 1L-MoS$_2$ sample connected to Au electrodes (yellow). The MoS$_2$ flake is marked by the dashed white line. The samples are covered by a few layer hBN flake marked by green dashed line. The left inset is showing the vertical structure of the sample schematically. Right: Raman spectrum of three different layer MoS$_2$ samples (monolayer, bilayer and 4-layers). The peak separation between $E_g$ and $A_{1g}$ are also shown. The excitation laser source was 532 nm.

To measure the VBM splitting directly, we have conducted photocurrent spectroscopy (PCS) at a varying temperature from 77K to 300 using a microscopy cryostat. We annealed every device using 532 nm high power laser (~200 mW) while the device was kept at 77 K inside the cryostat (see Methods for details). We illuminate devices using a low-intensity broadband white light from a thermal light source and record photocurrent generated from the device across a range of photon wavelengths (see Methods for details). In total, we studied 13 MoS$_2$ devices and 13 WS$_2$ devices (see Supporting Information for details).

Fig.2b presents the photocurrent spectroscopy for 1L- 5L MoS$_2$ samples measured at 77K. We observed clearly A, B, and C peaks in the spectrum as marked in Fig.2a. We also observed a small peak between B and C peaks, whose origin is not clear and beyond the scope of the paper. Further study is necessary to identify this new peak.

We present the photocurrent data in terms of photoresponsivity (i.e., the photocurrent generated per unit optical power). The sample was mounted inside a microscopy cryostat (Janis Research, ST-500) equipped with electrical feedthrough for electro-optical measurements. To collect the photocurrent, we biased the devices by different bias voltage ranging from 1V to 10V using a source-meter (Keithley 2400) (see Supporting Information for details).

Recently Zhang et al.,[13] have demonstrated the temperature-dependent properties of VBM in MoS$_2$ by using photoluminescence (PL) spectroscopy. Though the reported results are interesting, we argue that their conclusions are heavily affected by two characteristics of PL measurements. First, the B peak in PL measurements is not prominent compared to the A-peak, which makes it challenging to pinpoint the B peak energy, which is critical to determine the VBM splitting. Second, PL is a second-order process where a TMD first absorb the photon exciting electrons to a conduction band from the valence band and then reemits a photon as the electron transitions to the valence band. On the other hand, PCS is a more precise technique to determine the SOC than PL spectroscopy as the PCS provide equally strong A and B peaks and the spectrums follow the absorption spectrums. We measured properties of VBM in TMDs that are significantly different than the results reported by Zhang et al.[13] In particular, we have found strong temperature-dependent VBM in 1L- and 2L- TMDs, whereas Zhang et al. demonstrated temperature independent VBM for 1L- and 2L- MoS$_2$.

PC spectroscopy (PCS) is very similar to absorption spectroscopy and allows one to study single- and many-body electronic states, and valence band splitting in TMDs.[19] Unlike absorption spectroscopy, PCS can be easily measured for an electrically contacted microscopic device in a cryogenic environment, as the device itself acts as its own photodetector. The peak separation in PCS of A and B peaks directly measures the VBM splitting in TMDs. We used Lorentzian fitting to determine the positions and errors of A and B peaks. Fig.2b presents the energy-resolved photoresponsivity spectrum at a different temperature ranging from 77 K to 270 K. We see that A and B peaks are red shifting as we increase the temperature. Fig.2c and 2d present the peak positions for the A and B peaks at different temperatures for different layered samples, respectively. We calculated VBM splitting from the difference between A and B peaks as shown in Fig.2e. The errors were calculated using the error propagation method.

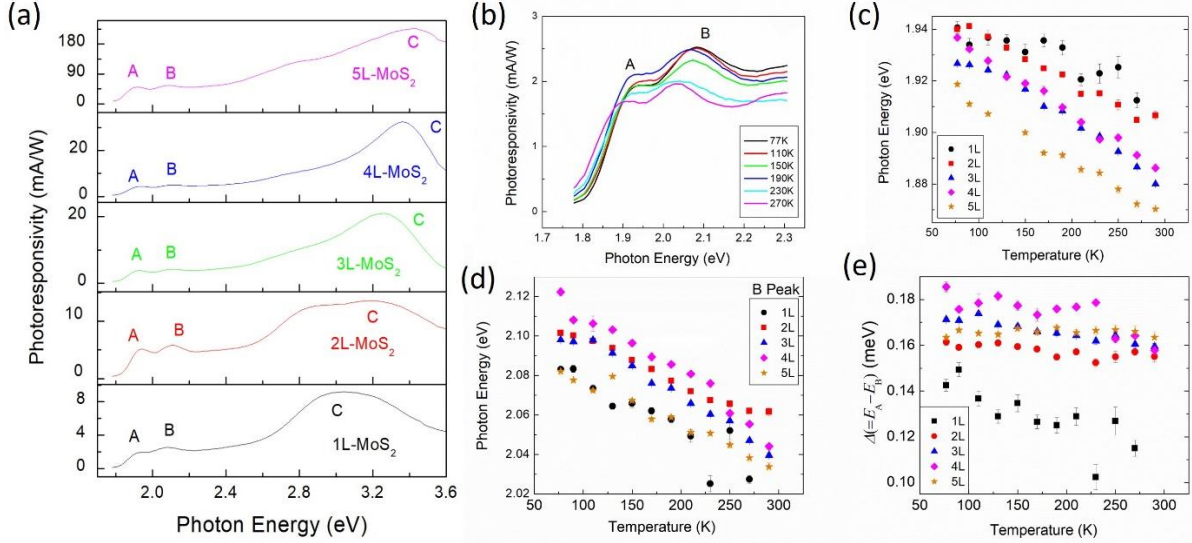

Figure 2: The VBM splitting of MoS$_2$. (a) Photocurrent spectra of different layered MoS$_2$ (1L to 5L). The left axis is presenting the photoresponsivity. (b) Photoresponsivity of a 1L-MoS$_2$ sample at different temperatures. (c) The plot is showing the A-peak position at different temperatures for different layered thickness. (d) The plot is showing the B-peak position at different temperatures for different layered thickness. (e) The difference between energy of the A and B peaks or VBM splitting $\Delta$ at different temperatures.

The temperature-dependent VBM coupling strength for different layered flakes reveals three important features. First, we observed that VBM splitting in 1L-MoS$_2$ strongly depends on temperature. This is very surprising as the current theory indicates that VBM splitting in 1L-TMDs originates from the spin-orbit coupling only, which should be temperature-independent as SOC is a relativistic effect. Second, the rate of change of the coupling strength ($m = \frac{\partial \Delta}{\partial T}$) depends on the layer thickness. It is the highest for a monolayer MoS$_2$ (-0.14 meV/K) and it reduces to vanishingly small for 5L-MoS$_2$. Interestingly, we found that the rate becomes positive (0.08 meV/K) for bulk (~100 layers) MoS$_2$. Third, the VBM splitting value is lowest for monolayer and increases as we increase the layer thickness.

To understand the effect of the transition-metal ions in VBM splitting, we also studied SOC coupling in WS$_2$ by using photocurrent spectroscopy at varying temperatures and layer numbers. We have studied samples of different thicknesses ranging from monolayer (1L) to five layers (5L). Fig.3a presents the photocurrent spectroscopy data recorded at 77K for 1L-5L samples.

Fig.3b shows the photocurrent spectra at different temperatures for a bilayer WS$_2$ sample. Interestingly, the photoresponsivity increases as we increase the temperatures. We attribute this effect to the reduction of resistance of the devices. Fig.3c and 3d present the peak positions of A and B peak at different temperatures, respectively. The peak positions and the corresponding errors were determined by fitting the Lorentzian function. Fig.3e presents the SOC coupling at different temperatures.

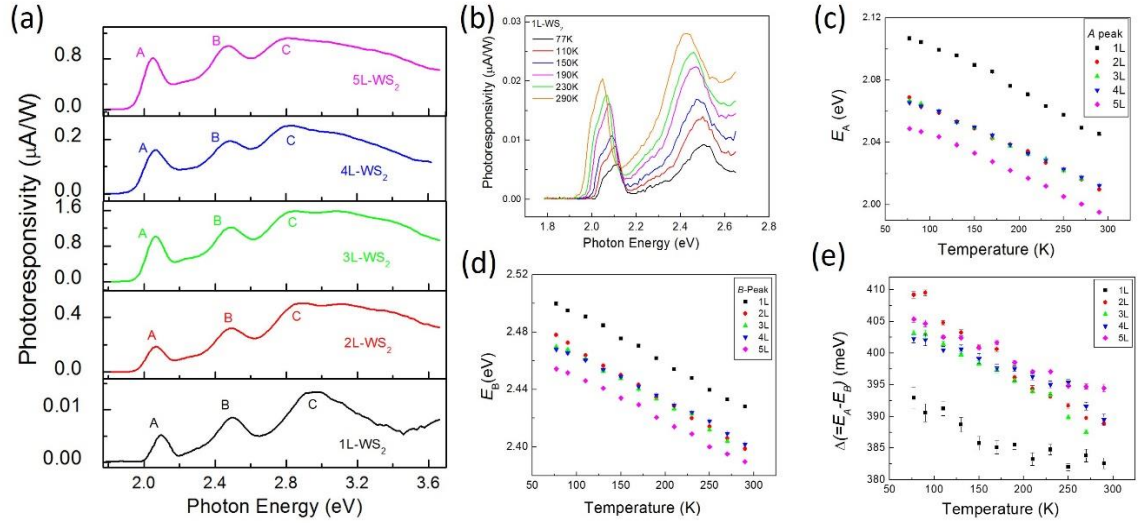

Figure 3: The VBM splitting of WS$_2$. (a) Photocurrent spectra of different layered WS2 (1L to 5L). The left axis is presenting the photoresponsivity. (b) Photoresponsivity of a 1L-WS2 sample at different temperatures. (c) The plot is showing the A-peak position at different temperatures for different layered thickness. (d) The plot is showing the B-peak position at different temperatures for different layered thickness. (e) The difference between energy of the A and B peaks, or VBM splitting $\Delta$ at different temperature for different thickness from 1L to 5L.

We observed 3 interesting features in the VBM splitting-temperature plots. First, VBM splitting, $\Delta$, reduces as we increase the temperature. The reduction of $\Delta$ in 1L-WS$_2$ confirms that the splitting is not originating solely from SOC in a 1L-WS$_2$ similar to the case in 1L-MoS$_2$ as shown in Fig.3. Second, the rate of change $\Delta$ with respect to temperatures remain independent of layer thickness, which different than the trend observed for MoS$_2$. Finally, the VBM splitting in monolayer WS$_2$ is the lowest and it increases for multilayer WS$_2$.

To quantify the temperature-dependent change of VBM, we calculated the rate of change $m = \frac{\partial \Delta}{\partial T}$, for all the samples. We also studied a bulk MoS$_2$ sample (layer number~100 determined by AFM thickness measurement) and a bulk WS$_2$ sample (layer number ~ 110). The calculated rate $m$ for all the samples studied in this experiment is shown in Fig.4. It is evident from the plot that the rate $m$ for MoS$_2$ increases as layer number increases up to 5L. Interestingly, we have not observed any clear trend for WS$_2$ samples depending on the layer thickness. For the bulk sample, $m$ becomes positive for MoS$_2$, whereas $m$ remains negative for WS$_2$. For two 10-layer MoS$_2$ samples, we observed widely separated $m$ values, whose origin is not clear to us.

It has been argued that the splitting can be entirely attributed to SOC at the monolayer-limit,[7-13] which suggests that the VBM splitting in 1L-TMDs should be independent of temperature. Instead, we observed the strongest temperature-dependent SOC for monolayer MoS$_2$ compared to a higher number of layers. Our result suggests that the VBM splitting at the monolayer limit is not solely due to SOC. We argue that the VBM splitting of 1L-MoS$_2$ originates from the mixing of SOC and the interlayer interaction with the substrate.

When a TMD is subjected to out-of-plane compressive pressure, it has been shown that the interlayer separation between adjacent TMD layer ($\delta$) decreases and the interlayer interaction between the layer increases.[18, 23] It has also been demonstrated that the Raman active vibrational modes (both $E_g$ and $A_{1g}$ modes) blueshift as the pressure is increased,[24] which suggests that the blueshift (redshift) of the Raman peak is a signature of the decrease (increase) of interlayer separation $\delta$ and increase (decrease) of interlayer interaction. Since an increase in temperature causes a redshift of both $E_g$ and $A_{1g}$ Raman peaks,[25] the average interlayer separation $\delta$ increases, and interlayer interaction decreases as one increases the temperature. Hence, we argue that the observed temperature-dependence of VBM splitting in TMDs is originating from the change in the interlayer separation that affects the interlayer coupling strength.

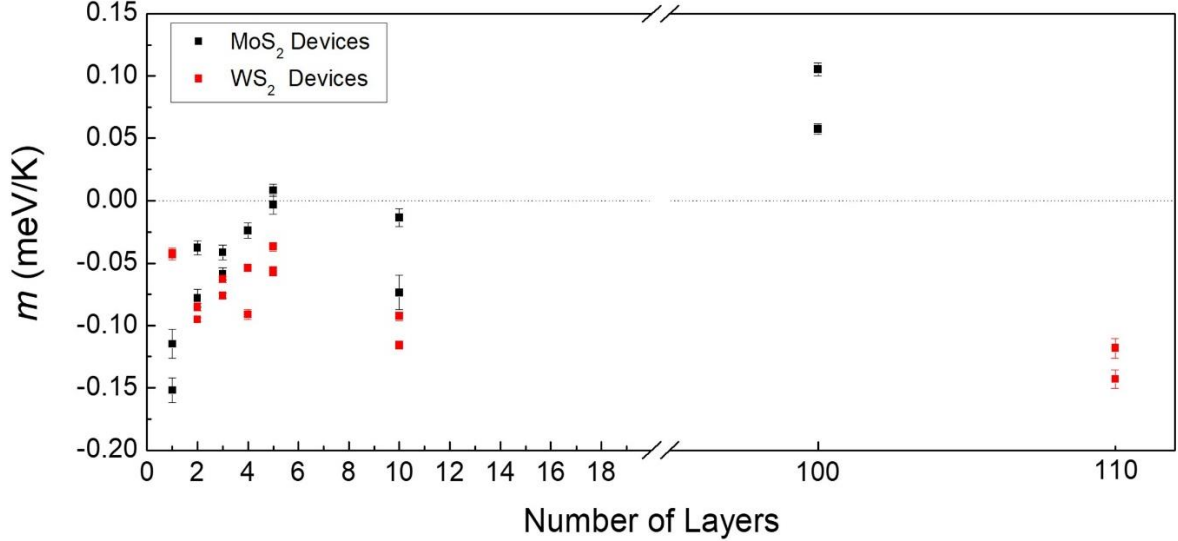

Figure 4: The rate ($m = \frac{\partial (E_A - E_B)}{\partial T}$) at which the splitting is behaving as we change the temperature for different layer MoS$_2$ and WS$_2$. To blow-up view of the splitting changing rate for 1L to 5L, a break in the x-axis is used. The error bars are also indicted.

To understand the effect of interlayer separation on VBM splitting, we conducted DFT and GW-BSE calculations to determine the band structure and absorption spectrums. In monolayer, there will be a combined effect of substrate dielectrics which is more pronounced in monolayer due to the reduced dimension and temperature effect. Moreover, the variability of the interface structure, effect of substrate strain, and dielectric screen effect with temperature is challenging and computational expensive that requires beyond DFT.[26, 27] That is why we conducted calculations for only a bilayer MoS$_2$ at varying interlayer separation.

We simulated the electronic band structure using the DFT calculations. The calculations were performed using the Vienna *ab-initio* simulation package (VASP).[28, 29] See Methods section for details. The electronic band structure calculated by DFT methods is shown in Fig.5(b), showing the VBM splitting at the K point of BZ.

To understand the excitonic A-B peak as a function of the interlayer separation of a bilayer MoS$_2$, we employ the GW-BSE method.[30] See Methods section for details. We observed from DFT calculations considering the relativistic effect that the VBM splitting is 155 meV in 2L MoS$_2$ for $d$(S-S)=3.08 Å. The calculated absorption spectra for three different values of $d$(S-S) = 3.08 Å, 3.33 Å, and 3.58 Å are shown in Fig.5(c). As we increase the interlayer separation, Δ (VBM) decreases as shown in Fig.5(d). To compare with the experimental observation, we have also re-plotted the Δ for a bilayer MoS$_2$ as shown in Fig.5(d).

Since we have measured Δ experimentally as a function of the temperature and calculated Δ as a function of interlayer separation, we can only make a qualitative comparison. Our results clearly demonstrate that the temperature dependence of Δ originates from the change in interlayer separation or interlayer interaction strengths. We have observed that Δ for a bilayer MoS$_2$ varies by ~ 10 meV as we change the temperature from

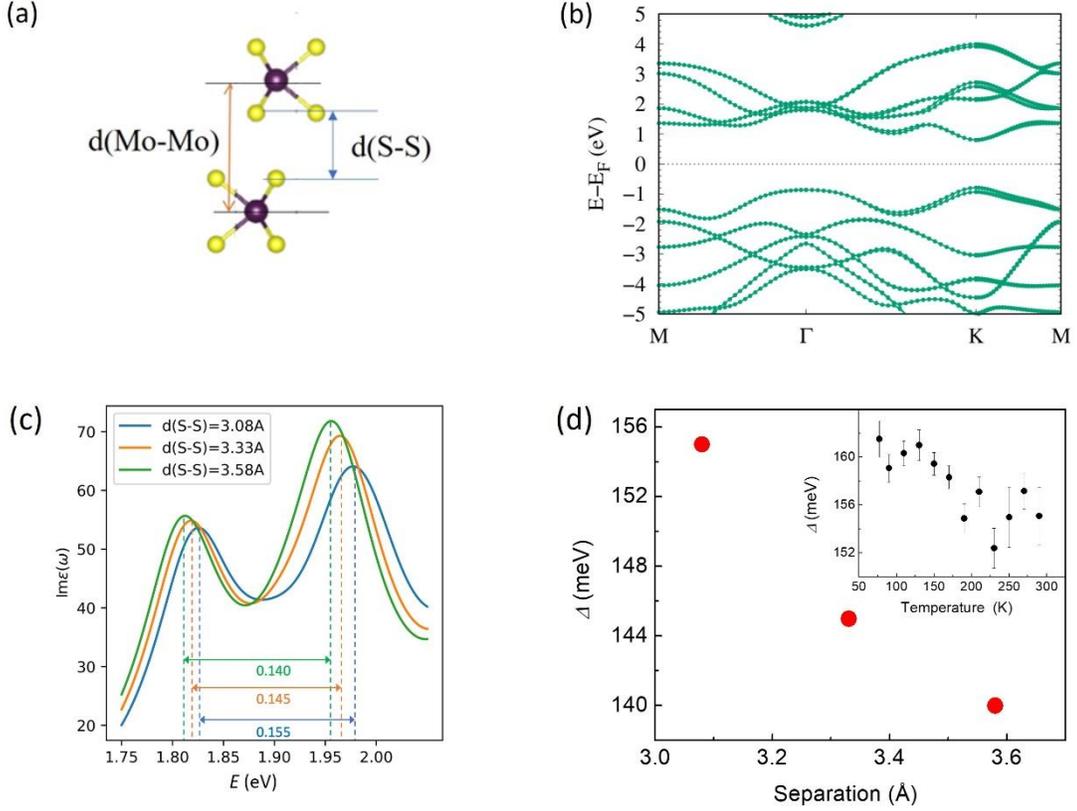

Figure 5: (a) Optimized atomic structure of a bilayer MoS$_2$. (b) Calculated electronic band structure with spin orbit coupling. (c) Absorption spectra for different interlayer $d$(S-S) separation for $d_0 = 3.08$ Å, $d_1 = 3.33$ Å, and $d_2 = 3.58$ Å. (d) Comparison of calculated VBM splitting values of a bilayer MoS$_2$ with the experimentally measured values. The main panel shows the calculated VBM splitting values for different interlayer separation. The inset is presenting experimentally measured VBM splitting for a 1L-MoS$_2$ as we change the temperature. The experimental results are the same data set presented in Fig.2(e).

77 K to 300 K. Interestingly, our simulation also predicts that the change in Δ very similar order as we increase the interlayer separation by 0.25 Å. We note that there is a ∼10 meV offset value between experimental measurements and simulations, but the trend still provides an explanation for the experimentally observed temperature-dependence of VBM splitting.

In conclusion, we studied temperature-, layer-, and material-dependent VBM for two different types of TMDs using photocurrent spectroscopy. We found that VBM depends on temperature, thickness, and materials of atomically thin TMDs. Our finding of VBM splitting in 1L-TMDs depending on temperature indicates that splitting Δ in 1L-TMDs does not originate solely from the SOC coupling. This result suggests that either SOC in 1L-TMDs is temperature-dependent or VBM splitting in 1L-TMDs is governed by the mixing of SOC and coupling strength to the substrate. Since SOC is a relativistic effect, we argue that VBM splitting in 1L-TMDs is caused by the mixing of SOC and the interaction with the substrate. We also found that the rate of change of SOC with respect to temperature is the highest for monolayer and the rate decreases as the layer number increases. To understand the effect, we have calculated the electronic band structure and VBM splitting for a bilayer with different interlayer separations, which suggests that the temperature-dependent VBM splitting in atomically 2L-TMDs can originate from the changes in the interlayer separation between neighboring layers.

Our study will help understand the intricate role spin-orbit coupling and interlayer interactions play in determining the VBM splitting in quantum materials.

**Acknowledgment**

G.B., V.Z.C., N. B., and A.K.M.N. acknowledge the support from the Department of Defense Award (ID: 72495RTREP). A.K.M.N. also acknowledges the support from the National Science Foundation Grant ECCS-1708907 and the faculty start-up grant provided by the College of Science and Engineering at San Francisco State University. S. KC acknowledges the faculty start-up grant provided by the Davidson College of Engineering at San Jose State University. Part of this research used resources of the National Energy Research Scientific Computing Center (NERSC), a U.S. Department of Energy Office of Science User Facility located at Lawrence Berkeley National Laboratory, operated under Contract No. DE-AC02-05CH11231 and Extreme Science and Engineering Discovery Environment (XSEDE), which is supported by National Science Foundation grant number ACI-1548562. T. S. and B. W. are supported by Basic Energy Sciences, Office of Science, Department of Energy (Grant No. DE-SC0020300). S.T acknowledges support from NSF CMMI-1933214, NSF DMR-2111812, CMMI-2129412, and NSF ECCS-2052527 as well as DOE-SC0020653. All AFM measurements were supported by NSF for instrumentation facilities (NSF MRI-CMMI 1626611). All Raman spectroscopy data were acquired at the Stanford Nano Shared Facilities (SNSF), supported by the National Science Foundation under award ECCS-2026822.

**Methods:**

**Sample Fabrication:** Atomically thin $MoS_2$ and $WS_2$ flakes were mechanically exfoliated from bulk crystals onto a heavily doped silicon substrate capped with a 90 nm thick thermal grown $SiO_2$ film. The $MoS_2$ flakes were obtained from naturally grown rock and $WS_2$ samples were grown by chemical vapor transport (CVT) technique. The number of TMD layers was characterized by using optical microscopy, Raman spectroscopy, and atomic force microscopy (AFM). For encapsulation with a few layers hBN, we prepare hBN flakes on $SiO_2$/Si substrate, which was picked by polyethylene terephthalate (PET). We used PET stamp to pick up the top hBN flake, atomically thin TMDs in sequence with accurate alignment using an optical microscope. The hBN/TMDs heterostructure was then stamped on a pre-fabricated Au electrode (70nm Au/ 5nm Cr) on a glass substrate. The patterned Au electrodes were fabricated using optical lithography followed by thermal evaporation of metals. All devices were prepared on a glass substrate to avoid the photogating effect.[19-21]

**Raman Characterization:** Confocal micro-Raman measurements were performed using commercial equipment (Horiba LabRAM Evolution). A 100× objective lens with a numerical aperture of 0.9 was used. The excitation source was a 532 nm laser (2.33 eV) with an optical power of ∼170 μW.

**Laser Assisted Thermal Annealing:** To obtain lower contact resistance, we annealed the devices using a 532 nm laser of beam power ~200 mW. The beam diameter is ~ 2 μm. We kept the devices at 77 K and under a bias voltage while we annealed the devices. The bias voltages were varied depending on the devices to obtain a measurable current. After annealing for 2 minutes, we measure the photocurrent spectroscopy. If the signal to noise ratio in photocurrent is low we repeated the laser assisted annealing process. We continued the process until we obtain a high signal to low noise ratio.

**Photocurrent Spectroscopy:** The photocurrent spectroscopy (PCS) at a varying temperature from 77K to 300 was conducted using a microscopy cryostat (Janis Research). We illuminate devices using a low-intensity broadband white light from a thermal light source and record photocurrent generated from the device across a range of photon wavelengths. The optical beam from the broadband thermal source (quartz halogen lamp) was directed through a monochromator (Acton Pro SP-2150i) and a mechanical chopper (45 Hz) onto the sample where it was focused down to a spot (~10 μm) with a diameter larger than the device. The photocurrent was measured by using a preamplifier (SRS570) connected to a lock-in-amplifier (SRS-830), which was locked to the chopping frequency. A commercial silicon photodetector (Hamamatsu S1223) was used to calibrate the light intensity incident on the sample.

**DFT calculations:** The DFT calculations were performed using the Vienna *ab-initio* simulation package (VASP).[28, 29] The nuclei and core electrons were described by the projector augmented wave function (PAW).[31, 32] The exchange-correlation used was the generalized gradient approximation (GGA) of Perdew–Burke–Ernzerhof (PBE) functional. All the structures were optimized until the maximum Hellmann-Feynman forces acting on each atom and the total energy is less than 0.01eV/Å and $10^{-5}$ eV, respectively. Each slab has a vacuum thickness of 20Å along z-direction to avoid the interaction due to periodic boundary conditions.

We used the GW-BSE method to calculate the absorption spectrum of TMDs.[30] The GPAW electronic structure calculations software,[33, 34] version 21.6.0, was employed to calculate the dielectric functions of the studied systems. Using the imaginary part of the dielectric function, optical adsorption spectrum and excitonic A and B peaks within the GW-BSE level of theory, which includes electron-hole correlations, were obtained. The reciprocal space was sampled by highly fine 30x30x1 k-points grid and truncated Coulomb interaction was considered to decouple the screening between periodic images.[35] The cut-off energy for the response function was set to 50 eV. The four highest energy bands below and the four lowest energy bands above the Fermi level were included in the GW-BSE calculations.

# Supporting Information

# Strong Effects of Interlayer Interaction on Valence-Band Splitting in Transition Metal Dichalcogenides

Garrett Benson[1], Viviane Zurdo Costa[1], Neal Border[1], Kentaro Yumigeta[2], Mark Blei[2], Sefaattin Tongay[2], K. Watanabe,[3] T. Taniguchi,[3] Andrew Ichimura[4], Santosh KC[5], Taha Salavati-fard,[6] Bin Wang,[6] and Akm Newaz[1]

[1]Department of Physics and Astronomy, San Francisco State University, San Francisco, California 94132, USA
[2] School for Engineering of Matter, Transport and Energy, Arizona State University, Tempe, Arizona 85287, United States;
[3] National Institute for Materials Science, Namiki 1-1, Tsukuba, Ibaraki 305-0044, Japan
[4]Department of Chemistry and Biochemistry, San Francisco State University, San Francisco, California 94132, USA
[5]Chemical and Materials Engineering, San Jose State University, San Jose, California 95112, USA
[6]School of Chemical, Biological and Materials Engineering, University of Oklahoma, Norman, Oklahoma 73019, United States

## 1) Pre-patterned Au electrodes Fabrication

To fabricate Au electrodes, we have selected are alkaline earth boro-aluminosilicate glasses obtained from Delta Technologies (C-1737). The glass thickness was 0.7 mm. The substrates were rinsed with copious deionized water, dried under argon, and then coated with positive photoresist (Microposit 1818). We have used 400 RPM spin coat rate to prepare a uniform photoresist film. After optical lithography, the wafer was transferred to a bell jar. We evacuated the bell jar to a base pressure of $10^{-7}$ Torr followed by a deposition Cr (5nm)/Au (100 nm) at a rate of 5 Å/s. Metal lift-off was conducted using an ultra-sonicator instrument. The fabrication process is shown in Fig.S1

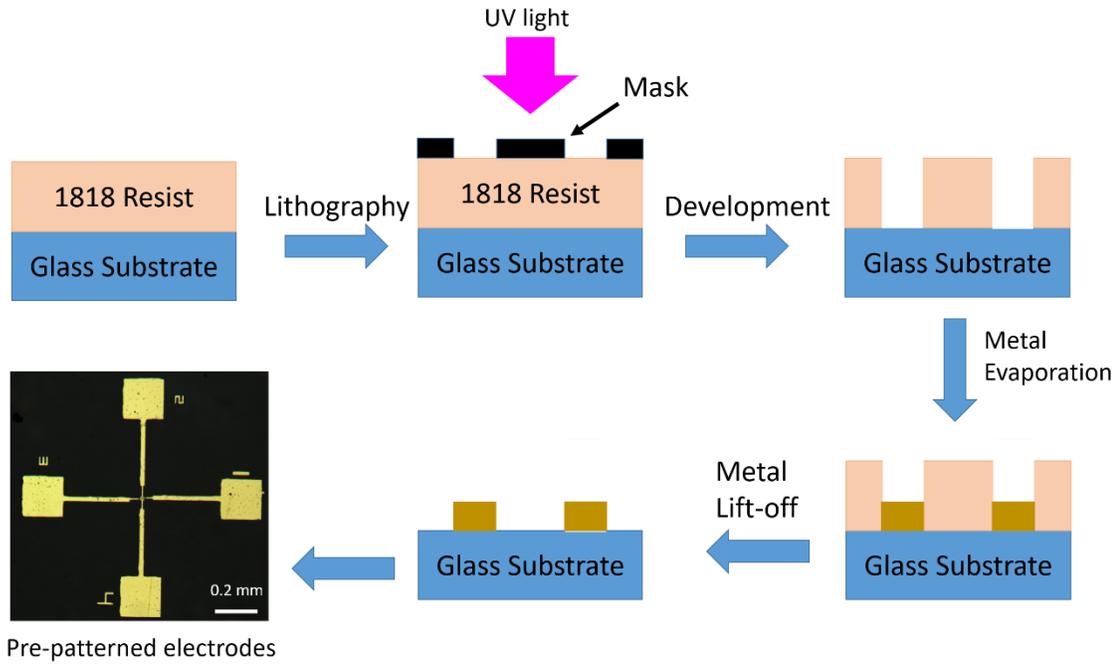

Figure S1: The fabrication process of the pre-patterned Cr/Au electrodes on a glass substrate. The optical image of a final fabricated electrodes on a boro-silicate glass is shown in the bottom left. The large metal squares (200 µm×200 µm) shown in the optical image are used for the wire bonding.

## 2) Monolayer WS2 sample characterization

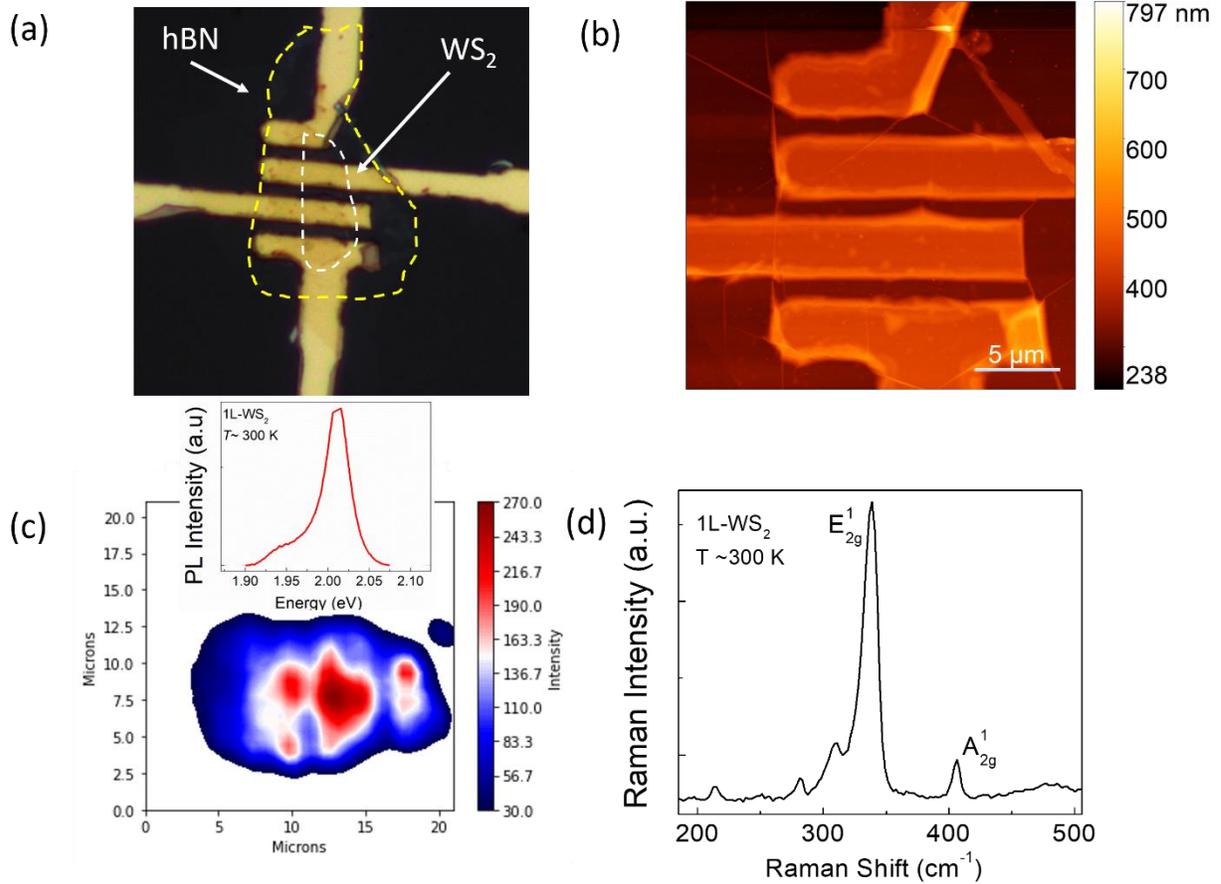

Figure S2: (a) Optical image of a 1L-WS$_2$ sample connected to Au electrodes (yellow). The WS$_2$ flake is marked by the dashed white line. The samples are covered by a few layer hBN flake marked by yellow dashed line. (b) An AFM image of the same sample. (c) The photoluminescence mapping of the sample at room temperature. This is a spatial contour plot showing the integrated PL intensity for different location of the laser beam on the sample. The inset presents the PL spectrum from the sample measured at room temperature. The laser excitation wavelength was 532 nm. (d) The Raman spectrum from the sample recorded at room temperature. The wavelength of the laser excitation was 532 nm.

## 3) Device information:

The table below presents the device information for MoS$_2$ and WS$_2$ samples.

| Device No. | Layer(s) | Bias Voltage (V) | PC Spectral Range (nm) | hBN |
|---|---|---|---|---|
| 9A | 1 | 6 | 540-700 | Yes |
| 9C | 1 | 8 | 300-700 | Yes |
| 1B | 2 | 2 | 300-700 | Yes |
| 11C | 2 | 4 | 300-700 | Yes |
| 13A | 3 | 4 | 540-700 | Yes |
| 13B | 3 | 3 | 340-700 | Yes |
| 5B | 4 | 2 | 300-700 | Yes |
| 15A | 5 | 3 | 340-700 | Yes |
| 15B | 5 | 1 | 540-700 | Yes |
| 27A | 10 | 4 | 580-680 | Yes |
| 27B | 10 | 3 | 340-700 | Yes |
| 17B | Bulk | 10 | 340-700 | No |
| 17C | Bulk | 10 | 340-700 | No |

**Table 1:** Information of MoS$_2$ devices. The second column presents the layer number. The bulk is used for devices that are ~100 layer thick. The third column presents the bias voltages used for photocurrent measurement. The fourth column presents the spectra range used for photocurrent measurement. The fifth column presents whether we cover the device the with atomically thin hBN flake.

| Device No. | Layer(s) | Bias Voltage (V) | PC Spectral Range (nm) | hBN |
|---|---|---|---|---|
| 33B | 1 | 1 | 340-700 | Yes |
| 33C | 1 | 1 | 340-700 | Yes |
| 18B | 2 | 6 | 540-700 | Yes |
| 18C | 2 | 8 | 300-700 | Yes |
| 23B | 3 | 2 | 300-700 | Yes |
| 23C | 3 | 4 | 300-700 | Yes |
| 21A | 4 | 4 | 540-700 | Yes |
| 21C | 4 | 2 | 300-700 | Yes |
| 22A | 5 | 3 | 340-700 | Yes |
| 22B | 5 | 1 | 540-700 | Yes |
| 22C | 5 | 4 | 580-680 | Yes |
| 28A | Bulk | 3 | 340-700 | Yes |
| 28B | Bulk | 10 | 340-700 | No |

0**Table 1:** Information of $WS_2$ devices. The second column presents the layer number. The bulk is used for devices that are ~100 layer thick. The third column presents the bias voltages used for photocurrent measurement. The fourth column presents the spectra range used for photocurrent measurement. The fifth column presents whether we cover the device the with atomically thin hBN flake.